\documentclass[conference]{IEEEtran}
\IEEEoverridecommandlockouts
\usepackage{cite}
\usepackage{amssymb,amsfonts}
\usepackage{graphicx}
\usepackage{textcomp}
\usepackage{xcolor}
\def\BibTeX{{\rm B\kern-.05em{\sc i\kern-.025em b}\kern-.08em
    T\kern-.1667em\lower.7ex\hbox{E}\kern-.125emX}}

\usepackage{helvet}
\usepackage{courier}
\usepackage[utf8]{inputenc}
\usepackage{array}
\usepackage{float}
\usepackage{booktabs}
\usepackage{multirow}
\usepackage{amsmath}
\usepackage{setspace}
\usepackage{subfig}
\usepackage{wasysym}
\usepackage{caption}
\usepackage{multirow}
\usepackage{algorithmic}
\usepackage[linesnumbered,ruled,lined,commentsnumbered]{algorithm2e}

 \usepackage{tgpagella}
\usepackage{times}
\PassOptionsToPackage{normalem}{ulem}
\usepackage{ulem}
\providecommand{\tabularnewline}{\\}
\usepackage{threeparttable}

\usepackage{enumitem}
\usepackage{hyperref}

\providecommand{\algorithmname}{Algorithm}

\floatname{algorithm}{\protect\algorithmname}
\SetKwComment{Comment}{\bgroup\hfill$\rhd$\~#1\egroup }{}

\newlength{\commentWidth}
\begin{document}

\title{Semantic-Enhanced Relational Metric Learning for Recommender Systems}

%
\author{
    \IEEEauthorblockN{Mingming Li$^{a,b}$, Fuqing Zhu$^{a,b}$, Feng Yuan$^c$, Songlin Hu $^a$}
    \IEEEauthorblockA{$^a$ Institute of Information Engineering, Chinese Academy of Sciences, China}
    \IEEEauthorblockA{$^b$ School of Cyber Security, University of Chinese Academy of Sciences, China}
     \IEEEauthorblockA{$^c$ School of Computer Science and Engineering, The University of New South Wales, Australia}
    \IEEEauthorblockA{\{{limingming},{zhufuqing}, {husonglin}\}@iie.ac.cn}  
}

 


\maketitle

\begin{abstract}
Recently, relational metric learning methods have been received great attention in recommendation community, which is inspired by the translation mechanism in knowledge graph. Different from the knowledge graph where the entity-to-entity relations are given in advance, historical interactions lack explicit relations between users and items in recommender systems. Currently, many researchers have succeeded in constructing the implicit relations to remit this issue. However, in previous work, the learning process of the induction function only depends on a single source of data (i.e., user-item interaction) in a supervised manner, resulting in the co-occurrence relation that is free of any semantic information. In this paper, to tackle the above problem in recommender systems, we propose a joint Semantic-Enhanced Relational Metric Learning (SERML) framework that incorporates the semantic information. Specifically, the semantic signal is first extracted from the target reviews containing abundant item features and personalized user preferences. A novel regression model is then designed via leveraging the extracted semantic signal to improve the discriminative ability of original relation-based training process. On four widely-used public datasets, experimental results demonstrate that SERML produces a competitive performance compared with several state-of-the-art methods in recommender systems.
\end{abstract}
\section{Introduction}

The recommender systems have become an indispensable tool in discovering personalized customer interests to alleviate information overload. The task is to recommend items that may be interest to users based on their historical records. Among all sorts of recommendation techniques during the past few decades, matrix factorization (MF)-based methods \cite{001koren2009matrix,002mnih2008probabilistic,mf-0salakhutdinov2008bayesian,BPR,mf-01zhang2013localized,liang2016factorization} are one of the most successful recommendation algorithms. MF-based methods predict user interests  using dot product calculation. However, the dot product suffers from metric limitations caused by the triangle inequality violation \cite{ram2012maximum}.

To satisfy such inequality property, a host of metric learning-based methods \cite{chen2012playlist,CML-hsieh2017collaborative,TransMF,RCL-tay2018latent} have been proposed to measure the similarity of user-item via Euclidean distance. Among them, relational metric learning methods are more competitive owing to the greater capability of modeling  a larger number of interactions, such as TransCF \cite{TransMF} and LRML \cite{RCL-tay2018latent}. The core idea of the relational metric learning methods is utilizing the latent relations as the translation operations between users and items, which is inspired by the successful adoption of translational metric learning in knowledge graph \cite{TransE}.
\begin{figure}[t]
\centerline{\includegraphics[scale=0.5]{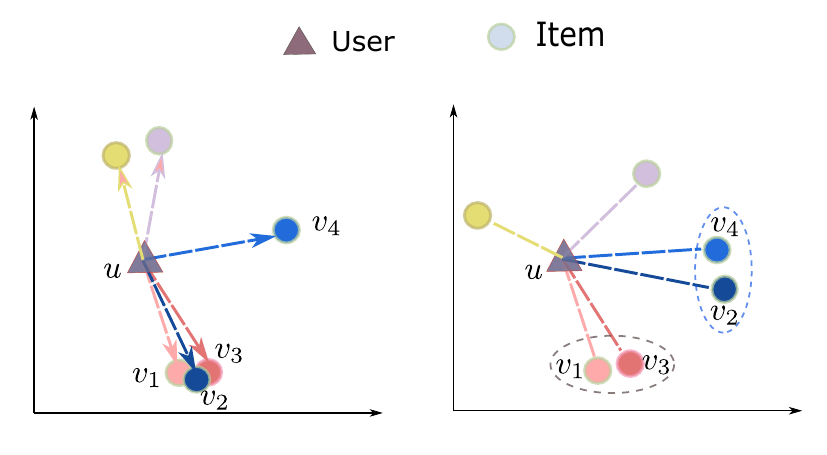}}
\caption{An illustrative example (left: LRML, right: the ideal case). For example,  $v_{1}$ and $v_{3}$
are similar items; $v_{2}$ and $v_{4}$ are similar items. }
\label{example}
\end{figure}

However, there is a significant difference between recommender systems and knowledge graph \cite{TransE,TransD,TransH,Rotate,Toruse}. In knowledge graph, the relations between entities are known in advance, while the similar relations between users and items are not given explicitly in recommender systems. Therefore, both TransCF and LRML directly construct the latent relations by an induction function and build the co-occurring relationship between users and items. Since the only source of the supervised signal comes from the user-item interaction matrix, the induced relations in the above methods fail to contain the \textbf{\textit{semantic information}} about users and items,  such as the \textbf{\textit{personalized interests of users}}. In other words, the optimization process is prone to overfitting, undermining the expression ability of the recommendation model. A concrete example is shown in Figure \ref{example} (left). There are four items $v_{1},v_{2},v_{3}$ and $v_{4}$ clicked by user $u$;  $v_{1},v_{2}$ and $v_3$ have the same historical records. We assume that $v_{1}$, $v_{3}$ are similar items, and $v_{2}$, $v_{4}$ are similar items. For  LRML   that only considers co-occurrences, similar interaction information leads to similar relationships, from which we can observe that $v_{1}$, $v_{2}$, and $v_{3}$ are located closely in the vector space and indistinguishable from each other, regardless of the differences between $v_1$ and $v_2$. An ideal case is shown in Figure  \ref{example} (right), each item could be adaptively translated to the user who clicks the current item according to the specific relationship of user with the item. Meanwhile, items of different categories could be far away from each other. For instance, $v_1$  is close to $v_3$, while $v_1$ stays away from $v_2$.

The above problem cannot be solved well by only relying on the user-item interactions. One method is introducing semantic signals into the model to guide the learning process of the relation induction, and thereby achieves a more excellent performance. \textbf{\textit{Which data can be used to extract the semantic signal?}} Among various choices, user target reviews have emerged as a promising candidate due to two advantages. First, target reviews are available and abundant in most datasets. Second, target reviews always contain the features of item and reflect the personalized interests of users, as they are written by the users to explain why they liked or disliked that particular item. \textbf{\textit{How does the signal guide the process of relation induction? }} We present a relational regression model to perform the supervised training, which makes the representation of a user-item relation close to that of its corresponding user reviews.

In this paper, we propose a novel joint learning framework named Semantic-Enhanced Relational Metric Learning (SERML) for recommender systems. The main idea is that the semantic information extracted from target reviews is set as the supervised signal for the learning process of relation induction. SERML consists of three modules: (1) textual representation learning module aims at extracting latent semantic information from the target reviews via hierarchical long short-term memory (HLSTM) and the attention mechanism; (2) relation induction module which is trained according to the additional semantic signal mentioned above for semantic-enhanced relation generation; (3) relational metric learning module which achieves the final triple modeling in accordance with users, items, and corresponding inductive relations. These three modules are incorporated into a joint framework for end-to-end training. Furthermore, we conduct a case analysis by tracking samples to prove our hypothesis. The main contributions of this paper are summarized as follows:
\begin{itemize}
\item We design a relational regression model that could generate a semantic-enhanced relation for each user-item pair;
\item We incorporate textual representation learning, relation induction, and relational metric learning into a unified framework, SERML, for end-to-end recommendation;
\item We conduct extensive experiments on four real-world datasets, and show that SERML produces a competitive performance compared with several state-of-the-art methods in recommender systems.
\end{itemize}

\section{Related Work}
This paper considers the formulation of relational metric learning based on the textual reviews in a unified end-to-end network. So we briefly review the methods of metric learning, textual representation learning and memory network.

\noindent \textbf{Metric Learning}. Briefly speaking, metric learning \cite{WeinbergerS09,ML-kulis2013metric,ML-1zadeh2016geometric,song2017parameter,do2019theoretically} aims to seek an appropriate distance function for input points, e.g., discrete distance, Euclidean distance, and Graph distance. In recommender systems, most of the researchers adopt the Euclidean distance to measure the similarity between a user and an item. In CML \cite{CML-hsieh2017collaborative}, the authors point out that the dot product-based measurement is limited, because it only considers the user-item relations, and presents a push mechanism by using metric learning.
FML \cite{FML2} also measures the distance between user and item via Euclidean distance. The above methods aim at measuring the positions of user vectors and item vectors in a metric vector space to make the particular preferred items of a user close to himself. Despite the success of metric learning, these methods suffer from an inherent limitation, called the ill-posed problem \cite{ill-pose}. To this end, the relational metric learning-based methods are proposed, which is inspired by the translation mechanism in knowledge graph \cite{TransE}. TransCF \cite{TransMF} constructs the user-item specific translation vectors from the neighborhood information of users and items, and then translates each user toward those items with which the user has relationships. LRML \cite{RCL-tay2018latent} induces the latent relations by memory-based attentive networks.

However, above latent relations are obtained by learning, making the optimization process tend to overfit. In addition, without specific semantic information, the inductive relations are more likely to represent user-item co-occurrences instead of the true interests of the users.

\noindent \textbf{Textual Representation Learning}. Textual representation learning has always been a hot topic in the natural language processing community and advanced in leaps and bounds recently. These technologies are widely applied in various downstream tasks, e.g., sentiment analysis \cite{amplayo2018cold,wang2018joint} and rating prediction \cite{TransNet-catherine2017transnets,DeepCONN3chen2018neural,DeepCONN4}. The most classical method \cite{012kim2014convolutional} uses convolutional neural network (CNN) to explore the textual information. Furthermore, researchers have managed to obtain the semantic representation of the document via HLSTM and the attention mechanism \cite{chen2016neural,lin2017structured}. In this paper, the main goal is to extract the personalized interests of users and features of items from target textual reviews. We employ the HLSTM and attention mechanism for textual representation learning.

\noindent \textbf{Memory Network}. The memory network was proposed in the context of question answering \cite{weston2014memory}. Two components are contained, i.e., an external memory (typically in the form of a matrix) and a controller (typically a neural network). The memory not only increases model capacity independent of the controller that operates on the memory, but also provides an internal representation of knowledge to track long-term dependencies. The memory network has not been applied to recommender systems until very recently \cite{ebesu2018collaborative,RCL-tay2018latent,zhou2019collaborative}. The success of the above models highlights that the memory network is effective and flexible to perform joint learning tasks.

\section{Methodology}

\label{ourmodel} In this section, we first describe the problem formulation, and then present a brief framework overview of SERML. Finally, we will give the detailed description of three modules in SERML, including textual representation learning, relation induction and relational metric learning.

\subsection{Problem Formulation}

In this paper, we formulate the top-N recommendation problem. There are a set of users $\mathcal{U}$ and a set of items $\mathcal{V}$. All the user-item interactions are noted as $\mathcal{I}=\left\{\left(u,v\right)|u\in\mathcal{U},v\in\mathcal{V}\right\}$. $\mathcal{N}_{u}\subseteq\mathcal{V}$ denotes the set of items that user $u$ has previously interacted. In addition, we collect additional information from historical records in which each user $u$ has given a review $T_{u,v}$ and an overall score $R_{u,v}$ for a particular item $v$. We denote all the reviews as $\mathcal{T}=\{T_{u,v}|u\in\mathcal{U},v\in\mathcal{V}\}$. Given the above information, we aim to predict whether a user $u$ has potential interests in an item $v\in\mathcal{V}\setminus\mathcal{N}_{u}$ with which he has interacted. In the rest of this paper, the bold uppercase letter is used to represent a matrix, while the bold lowercase letter is used to represent a vector. $\boldsymbol{\alpha}_{u}$, $\boldsymbol{\beta}_{v}$, and $\boldsymbol{r}_{u,v}$ denote user embedding, item embedding, and induced relation embedding, respectively.
\begin{figure*}[t]
\centerline{\includegraphics[width=13.5cm,height=6cm]{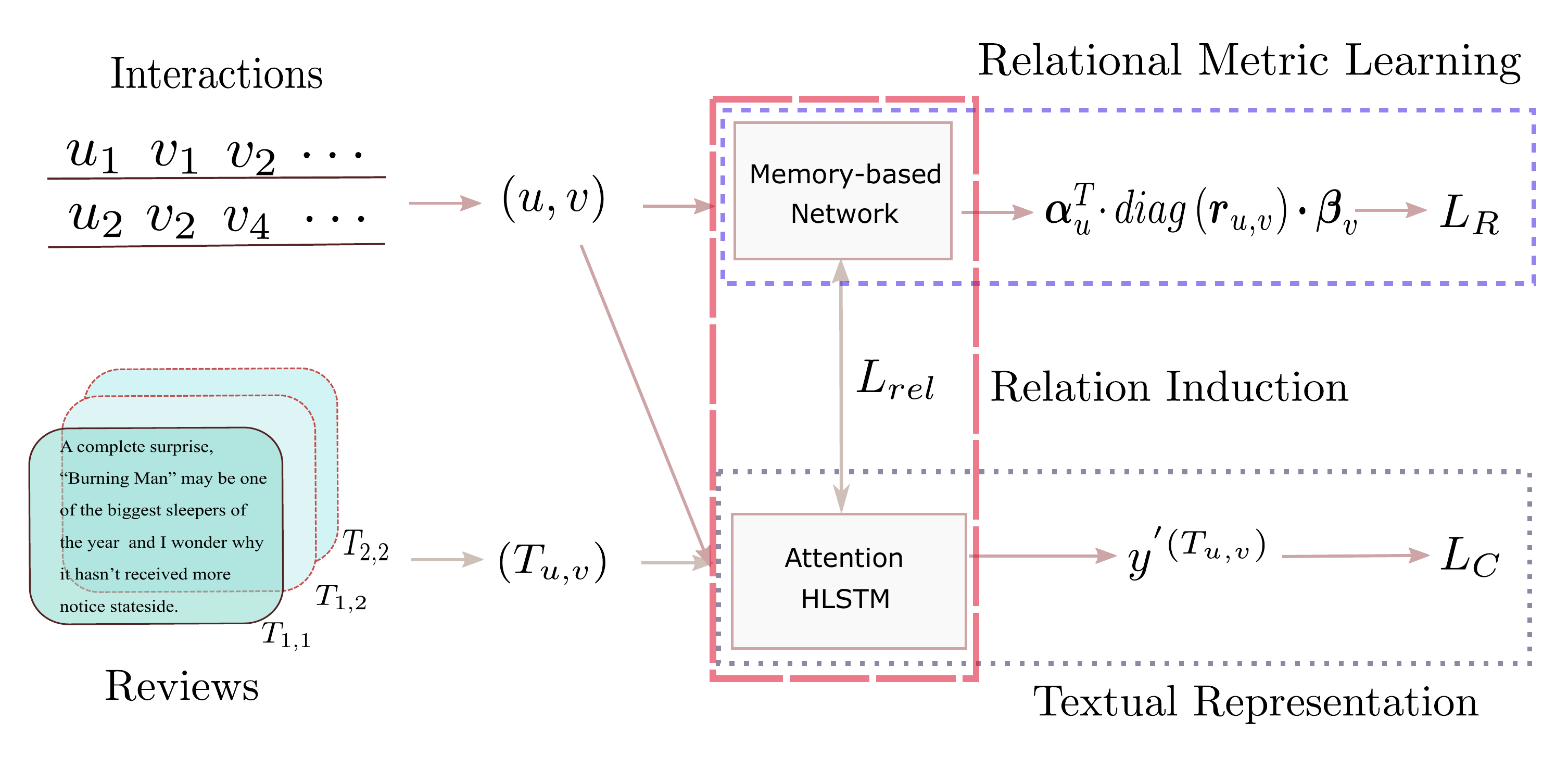}}
\caption{Framework illustration of SERML.}
\label{framework}
\end{figure*}
\subsection{Framework Overview}

As mentioned above, the primary motivation of SERML is exploiting extra semantic signals from the target reviews to guide the
learning process of user-item relation induction. Therefore, three modules are contained in our framework as shown in Figure \ref{framework}. Specifically,
\begin{enumerate}
\item The textual representation learning module aims at extracting latent semantic information via HLSTM and the attention mechanism.
\item The relation induction module generates semantic-enhanced relations by learning a relation induction function according to the supervised signal extracted from HLSTM.
\item The relational metric learning module models the triple interaction of users, items, and the corresponding inducted relations for pair-wise learning.
\end{enumerate}
Furthermore, we integrate the above three modules in an
end-to-end fashion through a joint framework. At last, we present
the prediction process on the test sets.

\subsection{Textual Representation Learning}

Given the target reviews of a user, the goal is to obtain the user representation in the vector space. To process the reviews, we have the following observations. Firstly, the same word or sentence may have different contributions to reveal the interest of the user across all the users and items in the system. Secondly, some words or sentences emphasize on the preferences of user, while some are more related to the features/aspectsof the item. Thus, we employ HLSTM and the attention mechanism on user and item information so as to capture various features on words and sentences.

\textbf{Input layer.} The input layer is the target reviews of user $u$ on item $v$, i.e., $T_{u,v}=\{\underset{s_{1}}{\underbrace{w_{1,1},\ldots,w_{1,|s_{1}|}}},\ldots,\underset{s_{L}}{\underbrace{w_{L,1},\ldots,w_{L,|s_{L}|}}}\}$, where $L$ is the number of sentences, $s_{l}$ denotes the $l$-th sentence, $|s_{l}|$ is the number of words in each sentence, and $K$ is the dimension of the word embedding.

\textbf{\textit{Word-level representation.}} Firstly, each input sentence $s_{l}=[w_{l,1},\ldots,w_{l,|s_{l}|}]$ is mapped to the embedding vector $[\boldsymbol{w}_{l,1},\ldots,\boldsymbol{w}_{l,|s_{l}|}]$, and the corresponding hidden states $[\boldsymbol{h}_{l,1}^{w},\ldots,\boldsymbol{h}_{l,|s_{l}|}^{w}]$ are generated by LSTM which could preserve the states over long periods of time. Secondly, user-specific words are extracted by applying the attention mechanism to the combined user and item information. Finally, we aggregate the representations of the above informative words to obtain the user-specific sentence representation ${\textstyle \boldsymbol{s}}_{l}$ that is formally expressed as follows:

\begin{alignat}{1}
e(\boldsymbol{h}_{l,j}^{w},\boldsymbol{\alpha}_{u},\boldsymbol{\beta}_{v}) & =\boldsymbol{W}_{1}^{w}\tanh(\boldsymbol{W}_{2}^{w}\boldsymbol{h}_{l,j}^{w}+\boldsymbol{W}_{3}^{w}[\boldsymbol{\alpha}_{u};\boldsymbol{\beta}_{v}]+\boldsymbol{b}_{w})\nonumber ,\\
 & c_{l,j}=\frac{\exp\left(e\left(\boldsymbol{h}_{l,j}^{w},\boldsymbol{\alpha}_{u},\boldsymbol{\beta}_{v}\right)\right)}{\sum_{k}^{|s_{l}|}\exp\left(e\left(\boldsymbol{h}_{k,j}^{w},\boldsymbol{\alpha}_{u},\boldsymbol{\beta}_{v}\right)\right)},\\
 & {\textstyle \boldsymbol{s}}_{l}=\sum_{j}^{|s_{l}|}c_{l,j}\boldsymbol{h}_{l,j}\nonumber,
\end{alignat}
where $c_{l,j}$ is the attention weight of $\boldsymbol{h}_{l,j}^{w}$ that measures the importance of the $j$-th word in the sentence $s_{l}$. $\boldsymbol{W}_{1}^{w}$, $\boldsymbol{W}_{2}^{w}$, and $\boldsymbol{W}_{3}^{w}$ are the weight matrices at the word level. $\boldsymbol{b}_{w}$ is the bias vector.

\textbf{\textit{Sentence-level representation.}} Similarly, given
the sequence of sentence vectors, $[\boldsymbol{s}_{1},\ldots,\boldsymbol{s}_{L}]$, the user-specific reviews can be represented as follows:

\textbf{
\begin{alignat}{1}
e(\boldsymbol{h}_{l}^{s},\boldsymbol{\alpha}_{u},\boldsymbol{\beta}_{v}) & =\boldsymbol{W}_{1}^{s}\tanh(\boldsymbol{W}_{2}^{s}\boldsymbol{h}_{l}^{s}+\boldsymbol{W}_{3}^{s}[\boldsymbol{\alpha}_{u};\boldsymbol{\beta}_{v}]+\boldsymbol{b}_{s})\nonumber, \\
 & c_{l}=\frac{\exp\left(e(\boldsymbol{h}_{l}^{s},\boldsymbol{\alpha}_{u},\boldsymbol{\beta}_{v})\right)}{\sum_{k}^{L}\exp\left(e(\boldsymbol{h}_{k}^{s},\boldsymbol{\alpha}_{u},\boldsymbol{\beta}_{v})\right)},\\
 & \boldsymbol{d}_{u,v}=\sum_{l}^{|L|}c_{l}\boldsymbol{h}_{l}^{s}\nonumber,
\end{alignat}
}where $c_{l}$ is the attention weight of $\boldsymbol{h}_{l}^{s}$.
$\boldsymbol{W}_{1}^{s}$, $\boldsymbol{W}_{2}^{s}$ and $\boldsymbol{W}_{3}^{s}$ are the weight matrices at the sentence level. $\boldsymbol{b}_{s}$ is the bias vector.

\textbf{Combination.} Normally, after obtaining the embedding $\boldsymbol{d}_{u,v}$ for the target reviews $T_{u,v}$, a softmax layer is then utilized to obtain a probability distribution $y^{'(T_{u,v})}$ over the sentiment classes (the sentiment classes are the corresponding rating scores, i.e., $C={1,2,\ldots,R^{max}}$). Finally, we apply the cross-entropy loss over all training documents $\mathcal{T}$ and arrive at the objective function $L_{C}$ as follows:

\begin{equation}
y^{'(T_{u,v})}=softmax\left(\boldsymbol{w}_{4}\boldsymbol{d}_{u,v}+\boldsymbol{b}_{c}\right),
\end{equation}

\begin{equation}
L_{C}=-\sum_{T_{u,v}\in\mathcal{T}}\sum_{c\in C}y_{c}^{T_{u,v}}\log\left(y_{c}^{'(T_{u,v})}\right),
\end{equation}
where $\boldsymbol{w}_{4}$ and $\boldsymbol{b}_{c}$ are the weight vector and inductive bias, respectively. $y^{(T_{u,v})}$ is the true probability distribution. All variables of the model are trained in an end-to-end manner. The optimization objective of sentiment classification is to minimize the loss $L_{C}$.

\subsection{Relation Induction}

As mentioned above, the historical records of user-item interactions lack explicit relations. Thus, a significant challenge is how to induce the relation vector between a user and an item. Without loss of generality, given a user-item pair, $\left(u,v\right)$, the induced relation can be defined as follows:

\begin{equation}
\boldsymbol{r}_{u,v}=f\left(\boldsymbol{\alpha}_{u},\boldsymbol{\beta}_{v}\right),
\end{equation}
where $f(\cdot,\cdot)$ is the induction function that aims to generate a specific relational vector. There are three candidate structures of the induction function: simple Element-wise multiplication\footnote{element-wise multiplication for the user and item.}, Multi-layer perception ($MLP$), and Memory-based network. In this paper, we adopt the Memory-based network to induce the user-item relations. For other structures, we also conduct extensive experiments for comparison in the following sections.

\subsubsection{Memory-based Network.}

Given a user-item pair, Memory-based network first applies the following steps to learn a joint embedding:

\begin{equation}
\boldsymbol{s}=\left[\boldsymbol{\alpha}_{u};\boldsymbol{\beta}_{v};\boldsymbol{\alpha}_{u}\odot\boldsymbol{\beta}_{v}\right],
\end{equation}
where $\odot$ is the element-wise operation. Next, we aim to learn an attention vector $\boldsymbol{\varpi}$, where the vector dimension is the same as $\boldsymbol{s}$. Then, we define a key matrix $\boldsymbol{K}$ , where $\boldsymbol{k}_{i}$ is the i-th row of $\boldsymbol{K}$. Each element of the attention vector $\boldsymbol{\varpi}$ can be defined as:

\begin{equation}
\varpi_{i}=\boldsymbol{s}^{T}\boldsymbol{k}_{i}.
\end{equation}
In order to normalize $\varpi_{i}$ to a probability distribution, we use the softmax function:

\begin{equation}
softmax(\varpi_{i})=\frac{e^{\varpi_{i}}}{\sum_{j}e^{\varpi_{j}}}.
\end{equation}
To generate the relation vector, we define a memory matrix $\boldsymbol{M}$ which can be interpreted as a store of conceptual building blocks ($\boldsymbol{m}_{i}$). Consequently, we use the above attention vector to generate a weighted representation of $\boldsymbol{M}$:

\begin{gather}
\boldsymbol{o}=\sum\varpi_{i}\boldsymbol{m}_{i},\\
\boldsymbol{r}_{u,v}=f\left(\boldsymbol{\alpha}_{u},\boldsymbol{\beta}_{v}\right)=\tanh(\boldsymbol{w}_{o}\boldsymbol{o}+\boldsymbol{b}_{o}),
\end{gather}
where $\boldsymbol{w}_{o}$ and $\boldsymbol{b}_{o}$ are the weight vector and the bias vector, respectively. The last transformation ensures that the induced relational vector is of the same dimension as $\boldsymbol{\alpha}_{u}$ and $\boldsymbol{\beta}_{v}$.

\subsubsection{Relation Regression.}

Importantly, the inductive relation $\boldsymbol{r}_{u,v}$ simply represents the abstract cross-feature of the user $u$ and item $v$. Consequently, items with the same interaction records tend to have similar relation vectors. In this sense, this strategy leads to a trivial solution that fails to distinguish the semantic differences among certain items. Therefore, we propose a novel relational regression model using the target reviews ($\boldsymbol{\widetilde{d}}_{u,v}=\boldsymbol{d}_{u,v}\boldsymbol{W}$) as the supervised signal to train the relation function. More concretely, we have:
\begin{alignat}{1}
 & \boldsymbol{\widetilde{d}}_{u,v}=\boldsymbol{r}_{u,v}+\boldsymbol{\epsilon}_{u,v},  \\ &
 \boldsymbol{\epsilon}_{u,v}\thicksim\mathcal{N}\left(\boldsymbol{0},\lambda^{-1}\boldsymbol{I}\right),
\end{alignat}
where $\boldsymbol{W}$ is the projection matrix that maps the embeddings of the target reviews into the same vector space as the user-item relationships; $\boldsymbol{\epsilon}_{*}$ is the vector offset, and $\boldsymbol{I}$ is the unit vector. We assume that the inductive relation $\boldsymbol{r}_{u,v}$ is close to the semantic vector but could diverge from it if it has to. Therefore, we can obtain $\boldsymbol{\widetilde{d}}_{u,v}\thicksim\mathcal{N}\left(\boldsymbol{r}_{u,v },\lambda^{-1}\boldsymbol{I}\right)$. The maximum likelihood function is:

\begin{equation}
\max\prod_{\left(u,v\right)\in\mathcal{I}}P\left(\boldsymbol{\widetilde{d}}_{u,v}|\mathcal{N}\left(\boldsymbol{r}_{u,v},\lambda^{-1}\boldsymbol{I}\right)\right).
\end{equation}
By further simplification, we have:

\begin{eqnarray}
 & \min-\log\prod_{\left(u,v\right)\in\mathcal{I}}P\left(\boldsymbol{\widetilde{d}}_{u,v}|\mathcal{N}\left(\boldsymbol{r}_{u,v},\lambda^{-1}\boldsymbol{I}\right)\right)\nonumber \\
 & \wasypropto\min\;\text{\ensuremath{\lambda}}\left\Vert \boldsymbol{\widetilde{d}}_{u,v}-\boldsymbol{r}_{u,v}\right\Vert _{2}^{2}\label{eq:regression}\\
 & \wasypropto\min\;L_{rel}=\text{\ensuremath{\lambda}}\left\Vert \boldsymbol{d}_{u,v}\boldsymbol{W}-f\left(\boldsymbol{\alpha}_{u},\boldsymbol{\beta}_{v}\right)\right\Vert _{2}^{2}\nonumber.
\end{eqnarray}
From the perspective of optimization, the objective function (\ref{eq:regression}) above is equivalent to a network of $f(\cdot,\cdot)$ with the semantic vector $\boldsymbol{\widetilde{d}}_{u,v}$ as the target.

\subsection{Relational Metric Learning}

Following the aforementioned procedure, given the interaction of $\left(u,v\right)$, a triple vector $\left(\boldsymbol{\alpha_{u}},\boldsymbol{r}_{u,v},\boldsymbol{\beta}_{v}\right)$ can be constructed. Then, we adopt a product-based function over the above triple to model the semantic symmetric relation. The score function is listed as follows:
\begin{equation}
s(\boldsymbol{\alpha}_{u},\boldsymbol{\beta}_{v})=\boldsymbol{\alpha}_{u}^{T}\cdotp diag\left(\boldsymbol{r}_{u,v}\right)\boldsymbol{\cdotp\beta}_{v}\label{score_function},
\end{equation}
where $diag(\boldsymbol{r}_{u,v})$ denotes a diagonal matrix whose diagonal elements equal to $\boldsymbol{r}_{u,v}$. Finally, the variables of all the models above are learned by minimizing a margin-based ranking objective which encourages the scores of positive relationships (triplets) to be higher than those of the negative relationships (triplets).
\begin{equation}
L_{R}=\sum_{\left(u,v\right)\in\mathcal{I}}\sum_{\left(u,v^{-}\right)\notin\mathcal{I}}\max\left[s\left(\boldsymbol{\alpha}_{u},\boldsymbol{\beta}_{v^{-}}\right)-s\left(\boldsymbol{\alpha}_{u},\boldsymbol{\beta}_{v}\right)+\xi,0\right],
\end{equation}
where $\xi$ is the margin. In this paper, we only use the push model \cite{song2017parameter}, because the push and pull model includes extra manually tuned parameters into the objective function, compared with the push model. These extra parameters serve as additional constraints that would make the searching space smaller \cite{WeinbergerS09} and the training process harder. Moreover, in TransCF, even with the extra penalty factor from the pull model included, the performance improves slightly on some datasets. Therefore, we select the push model due to its simlicity and strong enough constrained power.

\begin{table*}[t]
\caption{Statistics of four datasets. \label{tab:Statistics-of-dataset}}
\centering{}%
\begin{tabular}{c|ccccccc}
\toprule
\multirow{2}{*}{Datasets} & \multirow{2}{*}{\#users} & \multirow{2}{*}{\#items} & \multirow{2}{*}{\#ratings(\#docs)} & \multirow{2}{*}{\#docs/user} & \multirow{2}{*}{\#sens/doc} & \multirow{2}{*}{\#words/sen} & \multirow{2}{*}{sparsity}\tabularnewline
 &  &  &  &  &  &  & \tabularnewline
\midrule
Instant Video  & 4,902  & 1,683  &  36,486  & 7.24 & 6.23 &  15.84 & 0.9956\tabularnewline
Automotive & 2,788 & 1,835 & 20,218  & 6.99 & 5.79  & 15.91 & 0.9960\tabularnewline
Musical Instruments  & 1,397 & 900  & 10,216 & 7.18 & 6.21 & 15.72 &  0.9919\tabularnewline
Yelp13 & 1,631 & 1,633 & 78,966  & 48.42 & 10.89 &  17.38 & 0.9704\tabularnewline
\bottomrule
\end{tabular}
\end{table*}
\begin{table*}[t]
\caption{Performance comparisons on item ranking task in terms of NDCG@N, H@N
on four datasets.}
\label{tab:results_all} \centering{}%
\begin{tabular}{cccccccccc}
\toprule
\multirow{2}{*}{Methods } & NDCG@5  & NDCG@10  & H@5  & H@10  & \hphantom{} & NDCG@5  & NDCG@10  & H@5  & H@10 \tabularnewline
\cmidrule{2-10}
 & \multicolumn{4}{c}{Instant Video} &  & \multicolumn{4}{c}{Automotive}\tabularnewline
\midrule
BPR  & 0.7196  & 0.7396  & 0.6918  & 0.7198 &  & 0.7509  & 0.7669  & 0.7255  & 0.7555 \tabularnewline
MLP & 0.6924 & 0.6931 & 0.6367 & 0.6426 & & 0.7606 & 0.7625 & 0.7279 & 0.7347\tabularnewline
NeuMF  & 0.7322  & 0.7405  & 0.6903  & 0.7188  &  & 0.7611  & 0.7622  & 0.7347  & 0.7621 \tabularnewline
CML  & 0.7258  & 0.7308  & 0.6911  & 0.7118  &  &  {0.7785}  &  {0.7810}  & 0.7553  & 0.7636 \tabularnewline
FML  & 0.6713  & 0.6874  & 0.6728  & 0.7229  &  & 0.7762  & 0.7801  &  {0.7596}  &  {0.7682} \tabularnewline
TransCF  &  {0.7497}  &  {0.7570}  &  {0.7059}  & 0.7406  &  & 0.7730  & 0.7772  & 0.7421  & 0.7548 \tabularnewline
LRML  & 0.7364  & 0.7420  & 0.6908  & 0.7132  &  & 0.7763  & 0.7779  & 0.7396  & 0.7455 \tabularnewline
\textbf{SERML}  & \textbf{0.7638}  & \textbf{0.7700}  & \textbf{0.7264}  & \textbf{0.7501}  &  & \textbf{0.7863}  & \textbf{0.7922}  & \textbf{0.7675}  & \textbf{0.7846} \tabularnewline
\midrule
\multirow{1}{*}{} & \multicolumn{4}{c}{Musical Instruments } &  & \multicolumn{4}{c}{Yelp13}\tabularnewline
\midrule
BPR  & 0.3881  & 0.4075  & 0.4224  & 0.4800  &  & 0.8005  & 0.7793  & 0.5316  & 0.6433 \tabularnewline
MLP &  0.4771 & 0.4653 & 0.4608 & 0.4857 & & 0.6708 & 0.6828 & 0.4638 & 0.5678\tabularnewline
NeuMF  & 0.4595  & 0.4720  & 0.4498  & 0.4868  &  & 0.8004  & 0.7752  & 0.5431  & 0.6255 \tabularnewline
CML  & 0.4670  & 0.4748  & 0.4712  & 0.4917  &  &  {0.8237}  & 0.8000  & 0.5667  & 0.6653 \tabularnewline
FML  & 0.4261  & 0.4416  & 0.4546  & 0.5024  &  & 0.8094  & 0.7872  & 0.5495  & 0.6327 \tabularnewline
TransCF  & {0.4813} &  {0.4828}  &  {0.4839}  &  {0.5171}  &  & 0.8215  &  {0.8083}  &  {0.5675}  &  {0.6697} \tabularnewline
LRML  & 0.4709  & 0.4755  & 0.4634  & 0.4790  &  & 0.8196  & 0.8001  & 0.5630  & 0.6638 \tabularnewline
\textbf{SERML}  &  \textbf{0.4853}   & \textbf{0.4831}   & \textbf{0.4946}  & \textbf{0.5317}  &  & \textbf{0.8274}  & \textbf{0.8119}  & \textbf{0.5692}  & \textbf{0.6762} \tabularnewline
\bottomrule
\end{tabular}
\end{table*}
\subsection{Joint Learning}
To train the model in an end-to-end fashion, we integrate the above three modules in a joint learning framework shown in Figure \ref{framework}. The complete loss function is described as follows:

\begin{equation}
\min\;L=L_{C}+L_{R}+\gamma\cdotp L_{rel}+\varrho\cdot\left\Vert \boldsymbol{W}\right\Vert _{F}^{2},
\end{equation}
where $\gamma$ and $\varrho$ are all parameters, $\left\Vert \cdot\right\Vert _{F}$ is the Frobenius norm. In order to prevent overfitting, we apply the regularizers, $||\boldsymbol{u}_{*}||_{2}\leq1$, $||\boldsymbol{v}_{*}||_{2}\leq1$, at the end of each mini-batch \cite{RCL-tay2018latent,TransMF}.

\subsection{Prediction}

In the prediction process, given a user-item pair $(u,v)$, the inductive relation between them is first obtained via $\boldsymbol{r}_{u,v}=f(\boldsymbol{\alpha}_{u},\boldsymbol{\beta}_{v})$ which has already been trained under the semantic signal. Then, the final score of the given pair $(u,v)$ is computed by Equation (\ref{score_function}). Finally, we sort the scores and recommend the corresponding top-N items to target users.

\section{Experiments}

Since rating prediction and item ranking are usually investigated
separately with different evaluation metrics, we evaluate the pro-
posed approach on these two tasks individually.   Specifically,   the goal of items  ranking task  to  investigate the performance with metric learning based methods; the rating prediction task aims to  compare the performance with review-based methods.  In each experiment,   we first describe the datasets and baselines. Second, we elaborate on the experimental setup. Third, we show our experimental results in details. 
Afer that, we give the results of model analysis in itmes ranking task since it's  more meaningful for recommendation. 

\section{Item Ranking}
\subsection{Datasets}

We conduct experiments on the following public datasets: the Amazon Product Review\footnote{http://jmcauley.ucsd.edu/data/amazon/} and the Yelp Dataset Challenge in 2013 (Yelp13)\footnote{https://www.yelp.com/dataset/challenge}. These datasets all contain explicit feedback ratings on a scale of 1 to 5 and corresponding reviews, which have been widely-used in previous works \cite{DeepCONN4,DeepCONN5,tay2018multi}. For the Amazon dataset, we preprocess the data in a 5-core fashion (with at least five reviews for each user and item). In this paper, we use three different categories of the Amazon dataset (i.e., \textbf{Instant Video}, \textbf{Automotive}, and \textbf{Musical Instruments}). For the \textbf{Yelp13} dataset, we use a 10-core setting to provide a denser dataset for comparison as \cite{chen2016neural}. Table \ref{tab:Statistics-of-dataset} summarizes the satistics of the four datasets.

\subsection{Evaluation Protocols}

To evaluate the recommendation performance, we randomly split each dataset into training, validation, and testing sets with ratio 80\%:10\%:10\%. For computation simplicity, we follow the strategy in \cite{TransMF,RCL-tay2018latent}, and sample 500 items that have no interaction with the target user. To evaluate the ranking accuracy and quality, we adopt two widely-used metrics \cite{FML2,wang2019multi}: Normalized Discounted Cumulative Gain (NDCG@N) and Hit Ratio (H@N). All the evaluation metrics follow the implementation of a well-known open source recommendation project \footnote{http://mymedialite.net/index.html}.

\subsection{Compared Methods}

To evaluate the proposed SERML, we compare the results with the following state-of-the-art methods in recommender systems:

\textit{(a) Matrix factorization based method}:

 \textbf{BPR} \cite{BPR} is a classical pair-wise learning-to-rank method, whose optimization criterion
aims to maximize the differences between negative and positive samples.

\textit{(b) Deep learning based methods}:

\textbf{MLP} \cite{NeuMF} is a module of neural collaborative filtering method for implicit feedback, which only uses
the multi-layered preceptron (MLP) to model the features.
\textbf{NeuMF} \cite{NeuMF} is a competitive neural collaborative filtering method, which combines multi-layered
preceptron (MLP) with generalized matrix factorization (GML) and computes the ranking scores with a neural layer instead of dot product.

\textit{(c) Pure metric learning based methods}:

\textbf{CML} \cite{CML-hsieh2017collaborative} is a collaborative metric learning method for recommendation.
It assumes that users and items  explicit closeness could be measured by using Euclidean introduce that satisfies the inequality property.
\textbf{FML} \cite{FML2} is also a metric learning
method. This method firstly converts the preference into distance and then replaces
the dot product with Euclidean distance.

\textit{(d) Relational metric learning based methods}:

\textbf{TransCF} \cite{TransMF} is a collaborative translational metric learning method.  It
constructs user–item specific translation vectors by employing the neighborhood information of users and items, and translates each user toward items according to the user’s relationships with the items.
\textbf{LRML} \cite{RCL-tay2018latent} is latent relational metric learning method inspired by TransE \cite{TransE},
which  employ a augmented memory module to  induce the latent relation for each user-item interaction.

\subsection{Implementation Details}

We implement our method in Tensorflow, while for the baselines, we use the released codes provided in the original papers. We optimize our model with the Adam optimizer and tune the learning rate in \{0.10, 0.05, 0.01\} for different datasets. The embedding size is fixed to 100 for all the methods and the batch size is fixed at 512. We tune the parameters of textual representation according to \cite{textual_3wu2018improving}. For variables, all weights are initialized by uniform distributions of $[-0.01,0.01]$, and all latent vectors (such as $\boldsymbol{u},\boldsymbol{v}$) are initialized by Gaussian distributions with a variance of $0.01$ and a mean of $0.03$. Without specification, we show the results with $\gamma$ set to 1, $\varrho$ set to 0.01, and $\xi$ set to 0.5.

\begin{table*}[t]
\caption{Statistics of datasets \label{tab:Statistics-of-dataset}}
\centering  
\begin{tabular}{|c|ccccccccc|}
\hline
\multirow{2}{*}{{\small{}Datasets}} & \multirow{2}{*}{{\small{}\#users}} & \multirow{2}{*}{{\small{}\#items}} & \multirow{2}{*}{{\small{}\#ratings(\#docs)}} & \multirow{2}{*}{{\small{}\#docs/user}} & \multirow{2}{*}{{\small{}\#docs/item }} & \multicolumn{2}{c}{\textbf{\small{} \#}{\small{}max interactions }} & \multirow{2}{*}{{\small{}\#sens/doc}} & \multirow{2}{*}{{\small{}\#words/sen}} \tabularnewline
\cline{7-8}
 &  &  &  &  &  & {\small{}  user} & {\small{}item} &  &   \tabularnewline
\hline
{\small{}Instant Video } & {\small{}4,902 } & {\small{}1,683 } & {\small{} 36,486 } & {\small{}7.24} & {\small{}22.03} & {\small{}123} & {\small{}455} & {\small{}6.23} & {\small{} 15.84} \tabularnewline
{\small{}Automotive} & {\small{}2,788} & {\small{}1,835} & {\small{}20,218 } & {\small{}6.99} & {\small{}11.16} & {\small{}51} & {\small{}169} & {\small{}5.79 } & {\small{}15.91}  \tabularnewline
{\small{}Baby} & {\small{}17,177 } & {\small{}7,047} & {\small{}158,311} & {\small{}8.27} & {\small{}22.81} & {\small{}125} & {\small{}780} & {\small{}6.45 } & {\small{}16.43}  \tabularnewline
{\small{}Digital Music} & {\small{}5,426 } & {\small{}3,568} & {\small{}64,475} & {\small{}11.68} & {\small{}18.14} & {\small{}578} & {\small{}272} & {\small{}12.17} & {\small{} 17.38}  \tabularnewline
{\small{}Grocery} & {\small{}13,979 } & {\small{}8,711} & {\small{}149,434} & {\small{}10.30} & {\small{}17.36} & {\small{}204} & {\small{}742} & {\small{}6.57 } & {\small{}15.34}  \tabularnewline
{\small{}Health} & {\small{}34,850 } & {\small{}18,533} & {\small{}342,262} & {\small{}8.97} & {\small{}18.69} & {\small{}292} & {\small{}1089} & {\small{}6.44} & {\small{}15.88}  \tabularnewline
{\small{}Musical Instruments } & {\small{}1,397} & {\small{}900 } & {\small{}10,216} & {\small{}7.18} & {\small{}11.40} & {\small{}42} & {\small{}163} & {\small{}6.21} & {\small{}15.72} \tabularnewline
{\small{}Office Products } & {\small{}4,798} & {\small{}2,419} & {\small{}52,673 } & {\small{}10.86} & {\small{}22.01} & {\small{}94} & {\small{}311} & {\small{}9.25} & {\small{}16.24} \tabularnewline
{\small{}Patio} & {\small{}1,672 } & {\small{}962 } & {\small{}13,077 } & {\small{}7.87} & {\small{}13.80} & {\small{}66} & {\small{}296} & {\small{}9.83} & {\small{}17.26}  \tabularnewline
{\small{}Sports \& Outdoors } & {\small{}31,176} & {\small{}18,355 } & {\small{}293,306 } & {\small{}8.32} & {\small{}16.14} & {\small{}296} & {\small{}1042} & {\small{}6.10} & {\small{} 15.47}  \tabularnewline
{\small{}Tools \& Home} & {\small{}15,438} & {\small{}10,214 } & {\small{}133,414 } & {\small{}8.08} & {\small{}13.16} & {\small{}142} & {\small{}504} & {\small{}7.32} & {\small{}16.23} \tabularnewline
{\small{}Toys \& Games } & {\small{}17,692 } & {\small{}11,924 } & {\small{}166,180} & {\small{}8.634} & {\small{}14.06} & {\small{}550} & {\small{}309} & {\small{}6.76} & {\small{}16.01}  \tabularnewline
\hline
\end{tabular}{\small \par}
\end{table*}
\begin{table*}[t]
\caption{\label{tab:Comparisons-of-different}Comparisons of different methods
in terms of RMSE( the number of latent factor is 5). }
\centering 
\begin{tabular}{|c|ccccccccc|cc|}
\hline
\multirow{2}{*}{Datasets} & BMF & HFT & CTR & RMR & RBLT & TNET & FML & ALFM & \textbf{SERML} & \multicolumn{2}{c|}{Improvement(\%)}\tabularnewline
 & (a) & (b) & (c) & (d) & (e) & (f) & (g) & (h) & (j) & j vs. a &   j vs. h\tabularnewline
\hline
Instant Video  & 1.162  & 0.999 & 1.014 & 1.039  & 0.978 & 0.996 & 0.977 &  0.967 & \textbf{\small{}0.950} & 19.8  & 1.7\tabularnewline
Automotive & 1.032 & 0.968 & 1.016 & 0.997 & 0.924 & 0.918 & 0.896 & 0.885 & \textbf{\small{}0.867} & 15.6   & 1.8\tabularnewline
Baby & 1.359 & 1.112 & 1.144 & 1.178 & 1.122 & 1.110 & 1.097 & 1.076 & \textbf{\small{}1.053} & 29.8   & 2.3\tabularnewline
Digital Music & 1.093 & 0.918  & 0.921 & 0.960 & 0.918 & 1.061 & 0.923 & 0.920 & \textbf{\small{}0.900} & 18.5  & 2.0\tabularnewline
Grocery & 1.192 & 1.016 & 1.045 & 1.061 & 1.012 & 1.022 & 1.004 & 0.982  & \textbf{\small{}0.974} & 21.5   & 0.8\tabularnewline
Health & 1.263 & 1.073 & 1.105 & 1.135  & 1.070  & 1.114 & 1.052 & 1.042 & \textbf{\small{}1.030} & 23.3   & 1.2\tabularnewline
Musical Instruments  & 1.004 & 0.972 & 0.979 & 0.983 & 0.946  & 0.901 & 0.874 & 0.893 & \textbf{\small{}0.821} & 17.5   & 7.2\tabularnewline
Office Products  & 1.025 & 0.879 & 0.898 & 0.934 & 0.872 & 0.898 & 0.852 & 0.848  & \textbf{\small{}0.829} & 19.1   & 1.9\tabularnewline
Patio & 1.180 & 1.041 & 1.062 & 1.077 & 1.032 & 1.046  & 1.013 & 1.001  & \textbf{\small{}0.982} & 19.8  & 2.0\tabularnewline
Sports \& Outdoors  & 1.130 & 0.970 & 0.998 & 1.019 & 0.964 & 0.990 & 0.948 & 0.933  & \textbf{\small{}0.927 } & 18.6 &  0.6\tabularnewline
Tools \& Home & 1.168 & 1.013 & 1.047 & 1.090 & 1.011 & 1.041 & 1.023 & 0.974  & \textbf{\small{}0.965} & 19.5   & 0.9\tabularnewline
Toys \& Games  & 1.072 & 0.926 & 0.948 & 0.974  & 0.923 & 0.951 & 0.923 & 0.902  & \textbf{\small{}0.890 } & 16.2  & 1.2\tabularnewline
Average & 1.140 & 0.991 & 1.017 & 1.037 & 0.981 & 1.004 & 0.965 & 0.952 & \textbf{\small{}0.932} & 20.8   & 2.0\tabularnewline
\hline
\end{tabular}
\end{table*}

\subsection{Experimental Results}

The experimental results are summarized in Table \ref{tab:results_all}.

Generally, BPR performs worst on four datasets in terms of different metrics because the prediction scores are obtained through the dot product of user and item latent factors, leading to inadequate user representations \cite{ranking}. NeuMF makes a better performance than BPR, indicating that the multi-layer network is more effective than matrix factorizaiton.

Comparing with BPR and CML, we can observe that CML achieves a more excellent performance. Such improvement probably comes from the fact that CML makes the positions of items closer to the positions of users in the metric space, which is consistent with the previous work \cite{CML-hsieh2017collaborative}. This result also demonstrates that metric learning is simpler and more effective than traditional dot product-based methods.

Comparing with relational metric learning methods (i.e., LRML and TransCF) and pure metric learning methods (i.e., CML and FML), we can observe that the relational vectors between the users and items play an important role in the model performance improvement on several datasets.  

Comparing with LRML, a stable baseline, we can observe that SERML makes a substantial improvement. The results indicate that the semantic signal is useful, improving the representations of users/items.

\section{Rating Prediction}

\subsection{Datasets}

We conduct experiments on publicly available datasets Amazon\footnote{http://jmcauley.ucsd.edu/data/amazon/}. 
Table \ref{tab:Statistics-of-dataset} summarizes the statistics
on experimental datasets. These datasets all contain explicit feedback
ratings on a scale of 1 to 5 and corresponding reviews, which
have been widely used for rating prediction tasks
in previous studies \cite{DeepCONN4,DeepCONN5,DeepCONN3chen2018neural}.
In this paper, we used 12 categories and focus on the 5-core version,
with at least 5 reviews for each user and item.
\subsection{Evaluation Protocols}
Since our task is prediction rating for explicit feedback, we judge
the performance of our method based on the most popular and widely
adopted standard metrics used in recommender systems as Root Mean
Squared Error (RMSE), which measures the divergences between the predicted
rating and ground-truth rating. 

\subsection{Compared Methods}

We evaluate proposed method comparing to several state-of-the-art methods, such as 
\textbf{BMF},\cite{001koren2009matrix},\textbf{CTR} \cite{003wang2011collaborative},
\textbf{HFT }\cite{HFT-mcauley2013hidden}, 
\textbf{RMR} \cite{RMR-ling2014ratings}, \textbf{RBLT} \cite{RBLT-tan2016rating},  \textbf{TNET} \cite{TransNet-catherine2017transnets} , \textbf{ALFM} \cite{DeepCONN3chen2018neural}, and the metric learning based method \textbf{FML} \cite{FML2}.

\subsection{Implementation Details}
For a fair comparison, we make the same setup with the state-of-the-art
method \cite{DeepCONN3chen2018neural}. For each dataset, we randomly
split it into training, validation, and testing set with ratio 80:10:10.
Because the experimental dataset is consistent with  \cite{DeepCONN3chen2018neural},
the results of BMF, HFT, CTR, RMR, RBLT, TNET, and ALFM are also referenced
by this paper \cite{DeepCONN3chen2018neural}. For the metric learning
method of FML, we implement it by ourselves in Tensorflow, and the
initial value of all variable are the same as the original paper \cite{FML2}.
 Our method is training with Adam and the initial learning
rate is almost in $\{0.01,0.005,0.001\}$, since different datasets
have different data distributions. The batch size of our method is
set to 64.
\begin{figure*}[t]
\begin{centering}
\subfloat[Instant Video]{\includegraphics[scale=0.50]{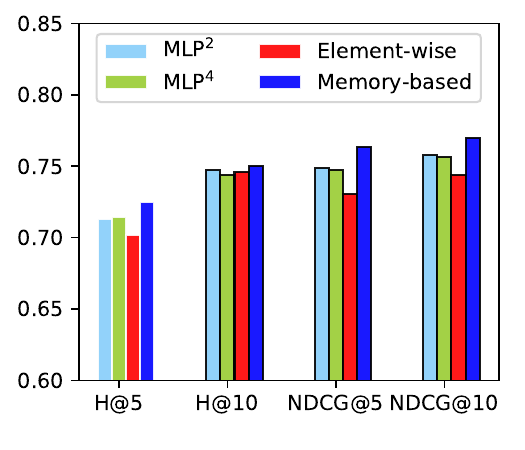}
}\subfloat[Musical Instruments]{\includegraphics[scale=0.50]{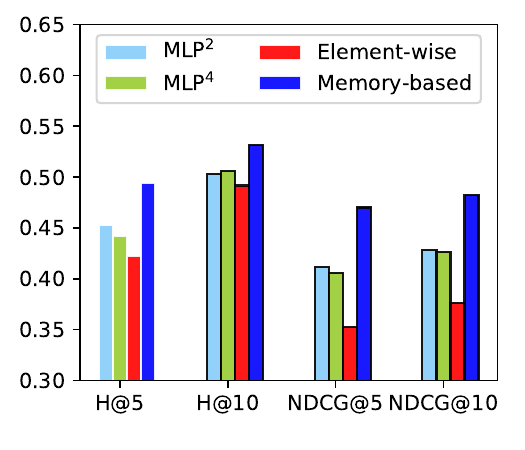}
}\subfloat[Automotive]{\includegraphics[scale=0.50]{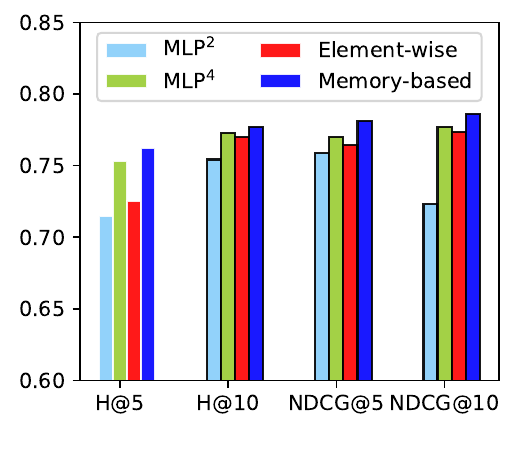}
}\subfloat[Yelp13]{\includegraphics[scale=0.50]{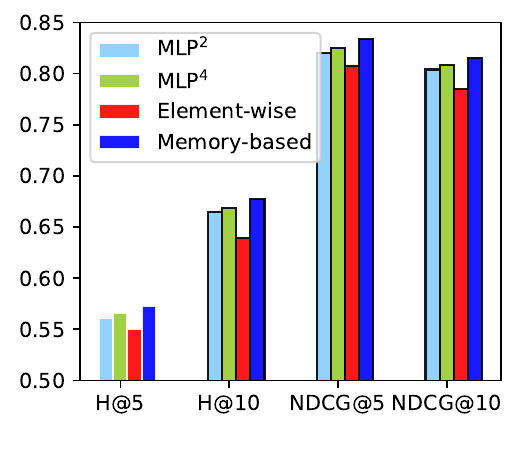}
}
\par\end{centering}
\caption{Experimental results of different induced relations: $MLP^{2}$, $MLP^{4}$,
Element-wise, and Memory-based.}
\label{multi-relations}
\end{figure*}
\subsection{Experimental Results}

The experimental results on twelve datasets are shown in Table \ref{tab:Comparisons-of-different}.
We can find that our proposed SERML consistently outperforms other
state-of-the-art methods on all datasets.

In general, these methods that both using rating and reviews information,
outperform those methods that only using rating information. These results expose the basic fact that
reviews information is quite useful and improves the representation
of user/item indeed. What's more, we also observe that review-based
methods, which learn the user/item embedding representation by utilizing
the deep learning, with greater performance than the topic model.
For example, TNET has a better improvement than HFT by 1.3\%. That
shows   deep learning based methods are effective to handle the natural
language processing problems. Besides, the aspect-based method also
makes good results by considering to capture the finer-grained interactions
between users and items at an aspect-level.

Importantly, FML makes a significant improvement than BMF and even
has better results than methods (such as ALFM, RMR) that use both
rating and reviews on most datasets. This also proven the metric learning
method is an useful metric tool than dot product, and it makes the
item position close to their user position by forcing them to satisfy
the conditions of triangle unequal.

Our proposed method not only considers the reviews information but
also uses the metric learning tools, which will have the advance of
both theoretically. The experimental results also verify the effective.
We also find that the performance of our
proposed is relational to the size of datasets.

\section{Model Analysis}
We evaluate SERML from the following perspectives: 1) Impact of different inductive strategies; 2) Impact of parameters ($\gamma$ and latent dimension); 3) Case analysis.
\subsubsection{Impact of Different Inductive Strategies}
 The relation induction module is essential in SERML, which aims to generate specific relations. Therefore, we vary the inductive strategy to investigate the impacts on the overall performance. As mentioned above, we conduct extensive experiments by using the $MLP^{2}$ (two-layer perceptron), $MLP^{4}$ (four-layer perceptron), and Element-wise multiplication to replace our memory-based network, respectively. The experimental results are plotted in Figure \ref{multi-relations}, in term of H@5, H@10, NDCG@5, and NDCG@10 on four datasets. We can observe that the memory-based attentive network outperforms the others, indicating that the memory-based network is effective and learns a weighted representation across multiple samples. Meanwhile, the noise is reduced and more informative features are selected for constructing the final inductive relations. Comparing with Element-wise multiplication and $MLP$, we can observe that $MLP$ tends to produce a more excellent performance, showing that the deep network could extract more abstract features to improve the generalization ability of the model.

\subsubsection{Impact of Parameters}
We investigate the rule of semantic signal for the relation induction by varying the value of $\gamma$. In this paper, we change $\gamma$ in the set of $\{0.001,0.01,0.1,1,10\}$. The experimental results on Amazon Instant Video and Yelp13 are shown in Table \ref{gama}. We can observe that SERML achieves the best performance on the Amazon Instant Video dataset when $\gamma=1.0$.

\begin{figure}
\centerline{
\subfloat[H@5]{\includegraphics[ scale=0.45]{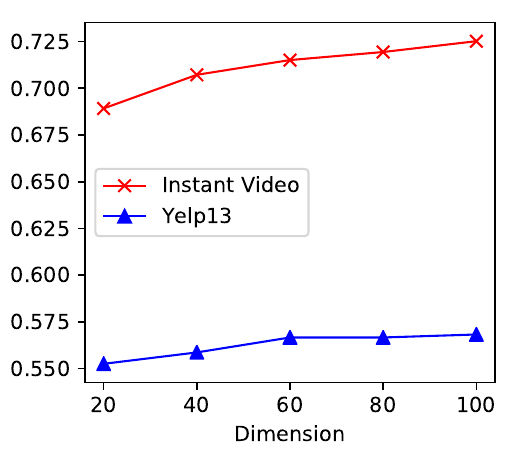}
}
\subfloat[NDCG@5]{\includegraphics[scale=0.45]{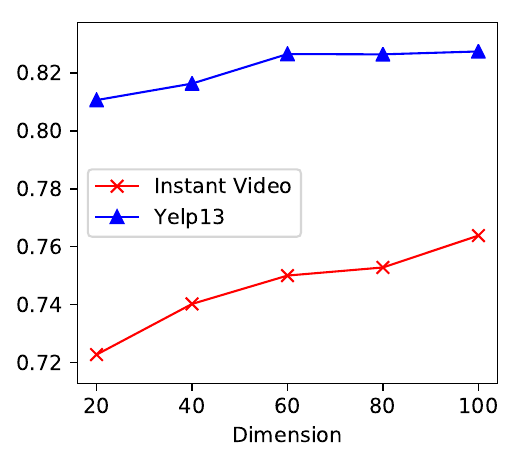}}
}
\caption{The effect of dimension on Instant Video and Yelp13.}\label{dimension}
\end{figure}

\begin{table*}[t]
\caption{Real samples data on Instant Video dataset.}\label{samples}
\centering{}%
\begin{tabular}{c|p{15cm}}
\toprule
{\footnotesize{}Rating} & {\footnotesize{}Target Review}\tabularnewline
\midrule
{\footnotesize{}$R_{1,1}=2.0$} & {\footnotesize{}$T_{1,1}$: What would you do on the eve of the \textcolor[rgb]{0.33,0.33,1.00}{ planet's
destruction}? The film has a \textcolor[rgb]{0.33,0.33,1.00}{pleasing grittiness} and  \textcolor[rgb]{0.73,0.33,0.83}{rawness}, and I
appreciated the use of actual news footage to \textcolor[rgb]{0.73,0.33,0.83}{make the apocalypse
seem tangible}.  Watched it with a friend, he \textcolor[rgb]{0.00,0.59,0.00}{bailed at the 30 minute}
mark... }\tabularnewline
\midrule
{\footnotesize{}$R_{1,2}=5.0$} & {\footnotesize{}$T_{1,2}$: It's almost \textcolor[rgb]{0.33,0.33,1.00}{infectious}. Anyone who \textcolor[rgb]{0.42,0.35,0.80}{loves
the movies} should not miss this one! Rarely do you get a chance to
sit down with so \textcolor[rgb]{0.33,0.33,1.00}{much talent and have them wax} on about the \textcolor[rgb]{0.00,0.55,0.55}{history
of film}. }\tabularnewline
\midrule

{\footnotesize{}$R_{1,3}=5.0$} & {\footnotesize{}$T_{1,3}$: An easy recommendation for
serious minded viewers, I \textcolor[rgb]{0.42,0.35,0.80}{loved} this movie. Teplitzky's \textcolor[rgb]{0.82,0.41,0.12}{``Burning
Man''} may easily be one of my \textcolor[rgb]{0.33,0.33,1.00}{favorite films} of the year, made all
the \textcolor[rgb]{0.00,0.55,0.55}{more surprising} in that I had no such expectations when I sat
down to \textcolor[rgb]{0.00,0.59,0.00}{watch} it. }\tabularnewline
\midrule
{\footnotesize{}$R_{1,4}=3.0$} & {\footnotesize{}$T_{1,4}$: \textquotedblleft America's Book of Secrets\textquotedblright{}
is one that is decidedly hit or miss. \textcolor[rgb]{0.42,0.35,0.80}{Playing up speculation, myth,
and even scandal}, the show aims to be somewhat provocative but, more
often the not, comes across as \textcolor[rgb]{0.00,0.59,0.00}{a bit silly}. The series never really qualified as
must-see \textcolor[rgb]{1.00,0.00,0.00}{ destination television} for me. }\tabularnewline
\bottomrule
\end{tabular}

\begin{tablenotes}
     \item[*] \footnotesize{The current userID is 'A27H9DOUGY9FOS'. And the itemID of $v_1$, $v_2$, $v_3$, and $v_4$ is 'B007OKD3GC', 'B0090EJZA8', 'B008NNY1U6', and 'B0071E6J30', respectively.} The genres of these items are different. Due to the limitations of space,
         only a few of each user's reviews are displayed.
   \end{tablenotes}
\end{table*}
\begin{table}
\caption{Impact of $\gamma$ for performance on Instant Video and Yelp13.}
\label{gama}
\centering{}{\footnotesize{}}%
\begin{tabular}{ccccc}
\toprule
\multirow{2}{*}{{\footnotesize{}\multirow{2}{*}{$\gamma$} }} & \multicolumn{4}{c}{{\footnotesize{}Instant Video}}\tabularnewline
\cmidrule{2-5}
 & {\footnotesize{}NDCG@5 } & {\footnotesize{}NDCG@10}& {\footnotesize{}\multirow{1}{*}{H@5} } & {\footnotesize{}H@10 } \tabularnewline
\midrule
{\footnotesize{}0.001 } & {\footnotesize{}0.7498 } & {\footnotesize{}0.7578}
& {\footnotesize{}0.7177 } & {\footnotesize{}0.7452 } \tabularnewline
{\footnotesize{}0.01 } & {\footnotesize{}0.7461 } & {\footnotesize{}0.7542} & {\footnotesize{}0.7124 } & {\footnotesize{}0.7420 } \tabularnewline
{\footnotesize{}0.1 } & {\footnotesize{}0.7520 } & {\footnotesize{}0.7607} & {\footnotesize{}0.7194 } & {\footnotesize{}0.7493 }
\tabularnewline
{\footnotesize{}1 }& \textbf{\footnotesize{}0.7638}{\footnotesize{} } & \textbf{\footnotesize{}0.7700} & \textbf{\footnotesize{}0.7264}{\footnotesize{} } & \textbf{\footnotesize{}0.7501}{\footnotesize{} } \tabularnewline
{\footnotesize{}10 } & {\footnotesize{}0.7225 } & {\footnotesize{}0.7340} & {\footnotesize{}0.6940 } & {\footnotesize{}0.7334 }
\tabularnewline
\midrule
\multirow{2}{*}{{\footnotesize{}\multirow{2}{*}{$\gamma$} }} & \multicolumn{4}{c}{{\footnotesize{}Yelp13}}\tabularnewline
\cmidrule{2-5}
 & {\footnotesize{}NDCG@5 } & {\footnotesize{}NDCG@10}& {\footnotesize{}H@5 } & {\footnotesize{}H@10 }\tabularnewline
\midrule
{\footnotesize{}0.001 } & \textbf{\footnotesize{}0.8276}{\footnotesize{} } & {\footnotesize{}0.8098} & {\footnotesize{}0.5679 } & {\footnotesize{}0.6695 } \tabularnewline
{\footnotesize{}0.01 } & {\footnotesize{}0.8241 } & {\footnotesize{}0.8076} & {\footnotesize{}0.5618 } & {\footnotesize{}0.6687 } \tabularnewline
{\footnotesize{}0.1 } & {\footnotesize{}0.8256 } & {\footnotesize{}0.8098} & {\footnotesize{}0.5668 } & {\footnotesize{}0.6709 } \tabularnewline
{\footnotesize{}1 } & {\footnotesize{}0.8274 } & \textbf{\footnotesize{}0.8119} & \textbf{\footnotesize{}0.5692}{\footnotesize{} } & \textbf{\footnotesize{}0.6762}{\footnotesize{} } \tabularnewline
{\footnotesize{}10 } & {\footnotesize{}0.8188 } & {\footnotesize{}0.8027} & {\footnotesize{}0.5606 } & {\footnotesize{}0.6637 } \tabularnewline
\bottomrule
\end{tabular}{\footnotesize \par}
\end{table}

We also investigate how the latent factor dimension affects the top-N recommendation  by varying [20,40,60,80,100]. Due to length limitation, we only show the results on Amazon Instant Video and Yelp13 datasets in Figure \ref{dimension}. We find that the performance is consistently improved with the increase of the latent dimension. Because the higher the dimension, the more potential features are encoded into the latent representation vectors.
 
\subsubsection{Case Analysis}
We make a simple example on the Amazon Instant video dataset to show the shortcomings of LRML and the advantages of SERML. First, we extract four items that have the same historical records from the training datasets. The corresponding ratings and target reviews are shown in Table \ref{samples}. According to LRML, we calculate the distances of the inductive relations between the extracted pairs via $\Delta_{v,k}^{r}=\left\Vert \boldsymbol{r}_{u,v}-\boldsymbol{r}_{u,k}\right\Vert _{2}^{2}$. Results are displayed as follows: $\varDelta_{1,3}^{r}=0.050$, $\varDelta_{2,3}^{r}=0.048$, $\varDelta_{1,2}^{r}=0.171$, $\varDelta_{14}^{r}=0.170$, and $\varDelta_{2,4}^{r}=0.445$. Similarly, we calculate the distances between items via $\Delta_{v,k}^{i}=\left\Vert \boldsymbol{\beta}_{v}-\boldsymbol{\beta}_{k}\right\Vert _{2}^{2}$. We obtain $\varDelta_{1,3}^{i}=0.52$, $\varDelta_{2,3}^{i}=0.49$, $\varDelta_{1,2}^{i}=1.085$, $\varDelta_{1,4}^{i}=1.084$, and $\varDelta_{2,4}^{i}=0.310$. We observe that $\boldsymbol{r}_{1,1}\thickapprox\boldsymbol{r}_{1,3}\thickapprox\boldsymbol{r}_{2,3}$ because of $\varDelta_{1,3}^{r}\thickapprox\varDelta_{2,3}^{r}$. In other words, $v_{1}$, $v_{2}$, and $v_{3}$ tend to have the same inductive relations. This phenomonon can be explained by the co-occurrence. Furthermore, we can also observe that $\varDelta_{1,3}^{i}\thickapprox\varDelta_{2,3}^{i}$, indicating that the final positions of $v_{1}$, $v_{2}$ are located in the neighborhood of $v_{3}$. Finally, according to the score function ($S_{v}^{u}=\left\Vert \boldsymbol{\alpha}_{u}+\boldsymbol{r}_{u,v}-\boldsymbol{\beta}_{v}\right\Vert _{2}^{2}$), the final sorted list is: $S_{3}^{u}>S_{1}^{u}>S_{4}^{u}>S_{2}^{u}$. However, the best sorted list is $S_{2}^{u}=S_{3}^{u}>S_{4}^{u}>S_{1}^{u}$. This result confirms that LRML converges into a suboptimal solution due to the lack of semantic information and interests of users. Similarly, we obtain the final relative positions of the above items as follows: $\varDelta_{1,2}^{i}=0.708$, $\varDelta_{1,3}^{i}=0.651$, $\varDelta_{1,4}^{i}=0.607$, $\varDelta_{2,3}^{i}=0.408$, and $\varDelta_{2,4}^{i}=0.518$. It is obvious that the items are scattered in the vector space according to the rule of different semantics. The final sorted list of our method in this case is: $S_{2}^{u}>S_{3}^{u}>S_{4}^{u}>S_{1}^{u}$, which is apparently more excellent than the ranking result of LRML.

\section{Conclusion}

In this paper, we point out that in metric learning-based recommendation methods, the induction function learned by the single source of user-item interactions tends to generate co-occurrence relations, without encoding personalized user preferences and item features. To remit this problem, we propose a joint learning framework SERML. Specifically, additional semantic information is extracted from the target textual reviews and then serves as the regressive signal for induction function learning. On four widely-used public datasets, experimental results demonstrate that SERML produces a competitive performance compared with several state-of-the-art methods.
\bibliographystyle{IEEEtran}
\bibliography{reference}

\end{document}